\documentclass[namedreferences]{SolarPhysics}
\usepackage{epsfig}

\usepackage{epsfig}

\newcommand{\be}{\begin{equation}}
\newcommand{\ee}{\end{equation}}
\newcommand{\beq}{\begin{eqnarray}}
\newcommand{\eeq}{\end{eqnarray}}

\begin{document}
\begin{article}
\begin{opening}

\title{ Prediction  Space Weather Using an Asymmetric Cone Model
for Halo CMEs}

\author{G. \surname{Michalek}$^{1}$,
        N.   \surname{Gopalswamy}$^{2}$,
        S. \surname{Yashiro}$^{3}$
       }

\runningauthor{MICHALEK ET AL.} \runningtitle{Prediction of space
weather }

\institute{$^{1}$ Astronomical Observatory of Jagiellonian
University, Cracow, Poland
                  \email{michalek@oa.uj.edu.pl}\\
           $^{2}$ Solar System Exploration Division, NASA GSFC,
Greenbelt, Maryland\\
           $^{3}$ Center for Solar and Space Weather, Catholic
University of America\\}

\date{Received ; accepted }

\begin{abstract}
Halo coronal mass ejections (HCMEs) are responsible of the most
severe geomagnetic storms. A prediction of their geoeffectiveness
and travel time to Earth's vicinity is crucial to forecast space
weather.
 Unfortunately coronagraphic observations are subjected to
 projection effects and do not provide true characteristics of CMEs.
Recently, Michalek~(2006, {\it Solar Phys.}, {\bf237}, 101)
developed an asymmetric cone model to obtain the space speed, width
and source location of HCMEs. We applied this technique to obtain
the parameters of all front-sided HCMEs observed by the SOHO/LASCO
experiment during a period  from the beginning of 2001 until the end
of 2002 ( solar cycle 23). These parameters were applied for the
space weather forecast. Our study determined that the space speeds
are strongly correlated with the travel times of HCMEs within
Earth's vicinity and with the magnitudes related to geomagnetic
disturbances.

\end{abstract}

\end{opening}
\section{Introduction}
Halo  coronal mass ejections (HCMEs) originating from regions close
to the central meridian of the Sun and directed toward  Earth cause
the most severe geomagnetic storms (Gopalswamy, Yashiro, and
Akiyama, 2007 and references therein). Therefore, it is very
important to determine the kinetic and geometric parameters
describing HCMEs. One of the most important parameter is the space
speed of CMEs used as input to CME and shock arrival models.
Unfortunately coronagraphic observations from the Sun-Earth line are
subjected to projection effects ({\it e.g.} Kahler, 1992; Webb, {\it
et al.}, 2000; St. Cyr {\it et al.}, 2000, Gopalswamy, Lara, and
Yashiro, 2003; Gopalswamy \emph{et al.}, 2001; Gopalswamy, 2004;
Gopalswamy, Yashiro, and Akiyama, 2007; Yashiro {\it et al.}, 2004).
There have been several attempts to obtain space speeds and other
parameters of CMEs (Zhao, Plunkett, and Liu (ZPL), 2002; Michalek,
Gopalswamy, and Yashiro (MGY), 2003; Xie, Ofman and Lawrence (XOL),
2004). These techniques
 need special measurements in the Large Angle Spectroscopic
 Coronagraph
 (LASCO; Brueckner {\it et al.}, 1995)
  field of view. These models assume that CMEs have cone
shapes and propagate with constant speeds. Recently, Michalek
(2006a) determined the space parameters of HCMEs with an asymmetric
cone model using the projected speeds obtained at different position
angles around the occulting disk. In the present study we use this
technique to get the space characteristics of all front-sided HCMEs
observed by LASCO in a period of time form  2001 until the end of
2002. Next, we use these parameters to obtain the travel times (TT)
of CMEs to Earth vicinity and the magnitudes  of the geomagnetic
disturbances ($D_{ST}$ index). The paper is organized as follows:
Section~2 describes the method used to determine the space
parameters presented here. In Section~3, we use the improved
parameters for space weather forecasting. Finally, conclusions are
presented in Section 4.

\section{Determination of the space parameters of HCMEs}
  Michalek (2006a) implemented a
cone model  to obtain the space parameters free from projection
effects. The model assumes that the shape of HCMEs is an asymmetric
cone and that they propagate with constant angular widths and
speeds, at least in their early phase of propagation. We can
determine the following HCME parameters: the longitude of the cone
axis ($\varphi$), the latitude of the cone axis ($\lambda$), the
angular width $\alpha$ (cone angle =$0.5\alpha$) and  the space
velocity $V_{space}$.  CMEs often have  a flux-rope geometry ({\it
e.g.}, Chen {\it et al.}, 1997; Dere {\it et al.}, 1999; Chen {\it
et al.}, 2000; Plunket {\it et al.}, 2000; Forbes 2000; Krall {\it
et al.}, 2001; Chen and Krall 2003), which encouraged us to
introduce the asymmetric cone model: the shape of CMEs
 is a cone but the cone cross section is an ellipse.
 The eccentricity and orientation of the ellipse are two additional parameters of the model. They are not
important for the space weather applications so we neglect them in
the present study. The following procedure was carried out to obtain
the parameters characterizing HCMEs. First, using the height-time
plots the projected speeds at different position angles (every
$15^{\circ}$) were determined.  This allowed us to obtain  $24$
projected velocities for a given HCME, which are required for the
fitting procedure. Second, using numerical simulation to minimize
the root mean square error, the cone model parameters were obtained.
Details of the numerical simulation and the equation used can be
found in Michalek (2006a). To save time, the simulation procedure
was performed with constraints on the cone model parameters. We
assumed that the space speed  is not smaller than the maximal
measured projected velocity for a given event. Second, using the
Extreme ultraviolet Image Telescope (EIT) (Delaboudini\'ere {\it et
al.}, 1995) and Solar Geophysical Data we determine the associated
eruptive phenomena (coronal dimmings, erupting filaments and
H$\alpha$ flares) which are coincident with the LASCO CME onset
time. This allows us to estimate source regions of HCMEs on the
solar disk and recognize front-sided events. The second assumption
on the cone model parameters is  that the cone model axis is
localized in a quadrant of the Sun where the associated phenomena
appear. To check these assumptions, for some events we performed the
simulation for a wider range of the cone model parameters.  Always,
the best fit cone model parameters fulfilled the above constraints.
Our numerical procedure allows us to place the apex of the cone at
the center of the Sun or on the solar surface.
 In the previous paper (Michalek, 2006a), we found
 that the better fits were
obtained when the apex of a cone is placed at the center of the Sun,
which we use   in this paper.

\section{Data}
The list of HCMEs studied in this paper is displayed in Table~1. We
considered only front-sided full HCMEs during the period of time
from the beginning of 2001 until the end of 2002.  We select this
limited period of time to get a representative sample of HCMEs which
could be use  to test our new cone model. In the SOHO/LASCO catalog
115 HCMEs are listed, 70 of which were front-sided. One of them was
too faint to perform necessary measurements. For the remaining 69
events height-time plots were obtained at different position angles
(every $15^{\circ}$). The projected speeds from the height-time
plots were then used for the fitting procedure to obtain the space
parameters of HCMEs. Using data from the World Data Center
(http://swdcdb.kugi.kyoto-u.ac.jp) geomagnetic disturbances caused
by these events were identified. In order to find a relationship
between HCMEs and magnetic disturbances a two step procedure was
performed. First, we found all geomagnetic disturbances, in the
considered period of time (2001-2002), with $D_{ST}$ index $\leq -30
nT$. This very high  limit ($-30 nT$) was chosen following Michalek
{\it et al.} (2006b). Such $D_{ST}$ values ($-30nT$) could occur
whether or not a CME  hits  Earth.  We assume that the associated
magnetic disturbance should start  no latter than 120 hours after
the first appearance of a given event in LASCO field of view and no
sooner than the necessary travel time of a given CME to Earth
calculated from the measured maximal projected velocity. We related
a given disturbance with a  HCME if they were within the specified
time range. Unfortunately we were not able to follow CMEs during
their entire trip to Earth, so there is some ambiguity in
associating the magnetic storms with CMEs.  During high solar
activity there are frequently more than one CME that could be
associated with a given magnetic disturbance. In our list there are
some magnetic disturbances associated with two different halo CMEs.
If we consider all CMEs included in the SOHO/LASCO catalog (not only
HCMEs) a number of multiple magnetic storms could be found.  Further
study into this association can be found in Gopalswamy, Yashiro, and
Akyama (2007).

20 events from our list were not geoeffective ($D_{ST}>-30nT$).
These HCMEs were slow or originated closer to the solar limb.   By
examining solar wind plasma data (from Solar Wind Experiment,
Ogilvie {\it et al.}, 1995) and interplanetary magnetic field data
(from Magnetic Field Investigation (Wind/MFI) instrument, Lepping
{\it et al.}, 1995), we identified interplanetary shocks driven by
respective interplanetary CMEs (ICMEs).   Measuring the time when a
HCME first appears in the LASCO's field of view and the arrival time
of the corresponding  shock at Earth  the travel time (TT) can be
determined (\emph{e.g.} Manoharan \emph{et al.,} 2004). The results
of our study are displayed  in Table~1. The
 first two columns are from the SOHO/LASCO catalog and give the date of the first appearance
 in the LASCO field of view and the
 projected speeds (V). The width and space speeds ($V_{space}$) estimated from the  cone model are shown in columns
 (3) and (4), respectively. In column (5) the r.m.s
error (in km~s$^{-1}$) for the best fits are given. The parameters
$\gamma$ and source locations are shown in columns (6) and (7),
respectively. In  column (8) the minimal values of $D_{ST}$ indices
for geomagnetic
  disturbances caused by  HCMEs are presented. Finally, in
  column (9) the travel time (TT)  of magnetic clouds to  Earth are given.

 \section{Implication for space weather forecast}
For the space weather forecast it is crucial to predict, with good
accuracy,  onsets (TT) and magnitudes ($D_{ST}$) of magnetic storms.
In the next two subsections, we consider these isues using the
determined space velocities.
\subsection{ Predictions of onsets  of geomagnetic  disturbances}
 Figure~1 shows the
scatter plots of the plane of the sky speeds (from SOHO/LASCO
catalog) versus the travel times. Diamond symbols represent events
originating from the western hemisphere and cross symbols represent
events originating from the eastern hemisphere. The dashed line is a
polynomial fit to data points (it is use third degree polynomial
function). Correlation coefficients are: 0.68 for the western and
0.49 for the eastern events. The standard error in determination of
the travel time (TT) is  $\pm$16 hours.

\begin{figure}    %%%%%%%%%%%%%%%%%% FIGURE 1
\begin{center}
\hbox{
\psfig{file=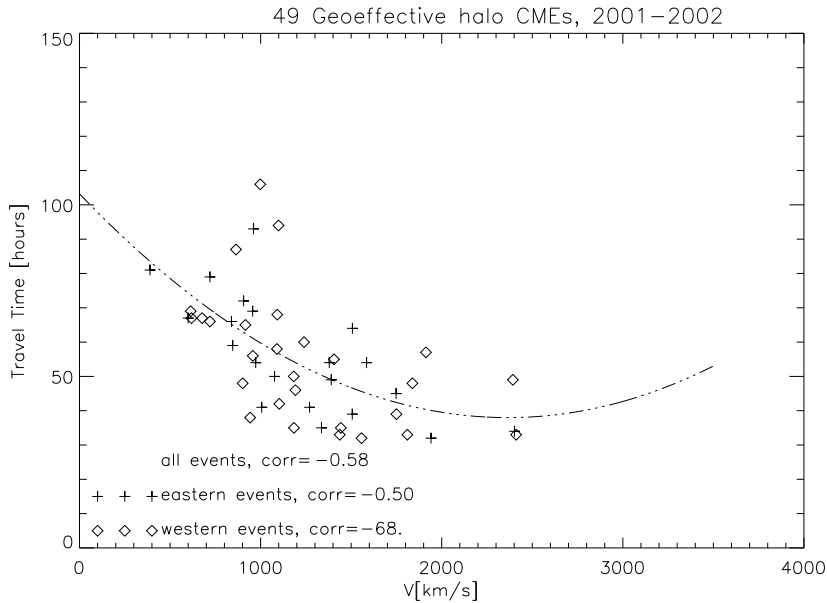,width=11.5cm,clip=}}

\caption{The scatter plot of the sky-plane speed versus the HCME
travel time  (TT). Diamond and cross symbols represent events
originating from the western  and eastern hemispheres, respectively.
The dot-dashed line is a polynomial fit to all the data points.}
\end{center}
\end{figure}
For comparison,   we present in Figure~2 (left panel) similar plot
except for the space speeds. The figure clearly shows that the space
speeds are strongly correlated with the TT. Now the correlations
coefficients are more significant:  0.71 for the western and  0.75
for the eastern events. The standard error in determination of the
travel time is only $\pm$10 hours. In  Figure~2 (right panel) we
show also similar plot but for the space speeds projected in the
Earth direction. To illustrate that our considerations are
consistent with previous results we compare them with the  ESA model
(the continuous line, Gopalswamy \emph{et al.} 2005b). For these
plots we used only the 49 geoeffective ($D_{ST}\leq-30nT$) events.
For comparison, in  Figure~2 (right panel) we added the three events
(2000/07/14, 2003/10/28, 2003/10/29) of historical importance,
represented by the dark diamonds.
%\begin{figure}    %%%%%%%%%%%%%%%%%% FIGURE 1
%\begin{center}
%\hbox{
%\psfig{file=figure1a.eps,width=7.cm,clip=}
%\psfig{file=figure1b.eps,width=5.cm,clip=}}
%\psfig{file=fig3new.ps,width=4.5cm,clip=}}
%\psfig{file=fig3new.ps,width=4.5cm,clip=}}
\begin{figure*}

\vspace{4.0cm} \includegraphics{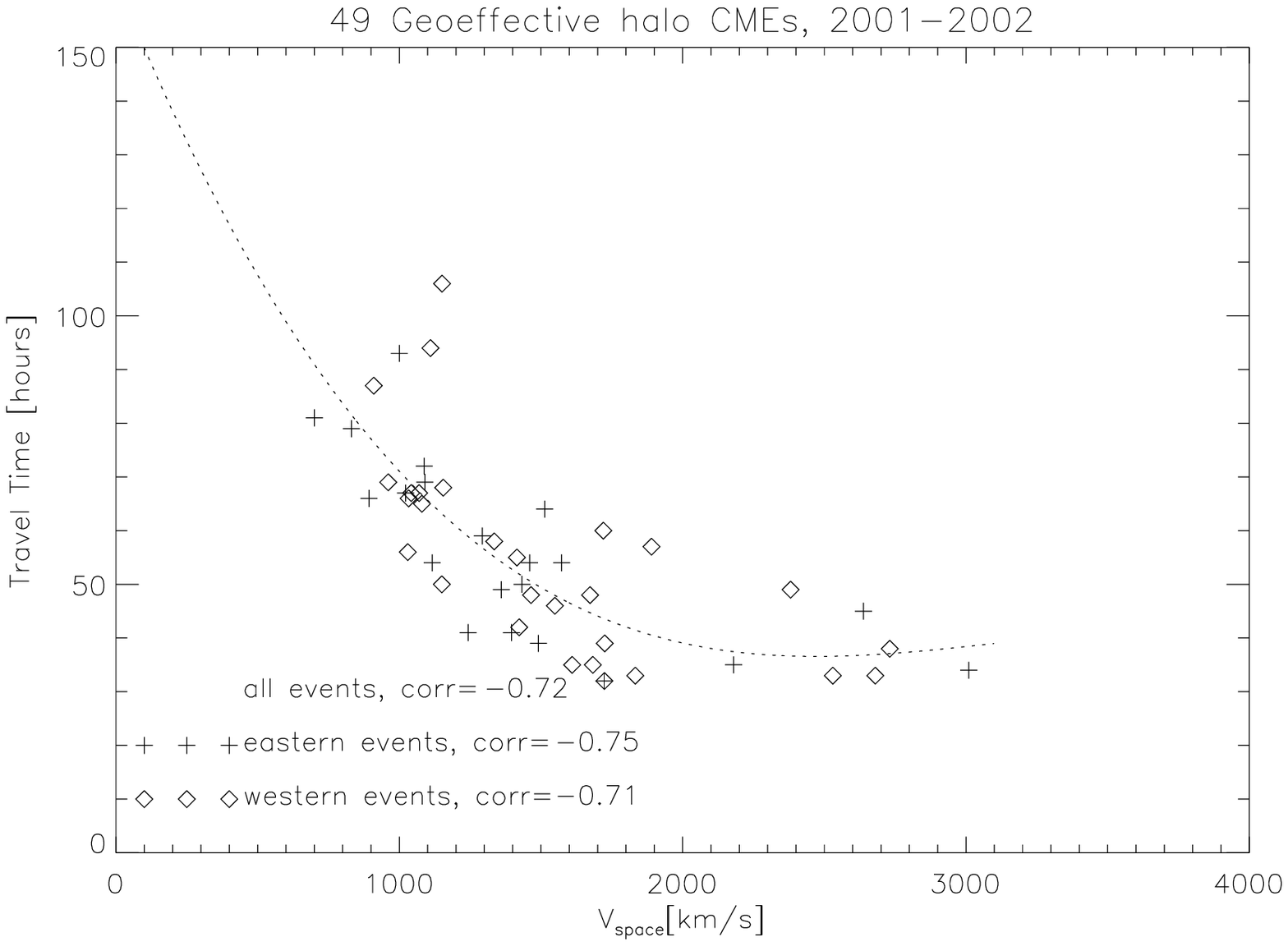} \vspace{3mm}

 \vspace{3.0cm}  \includegraphics{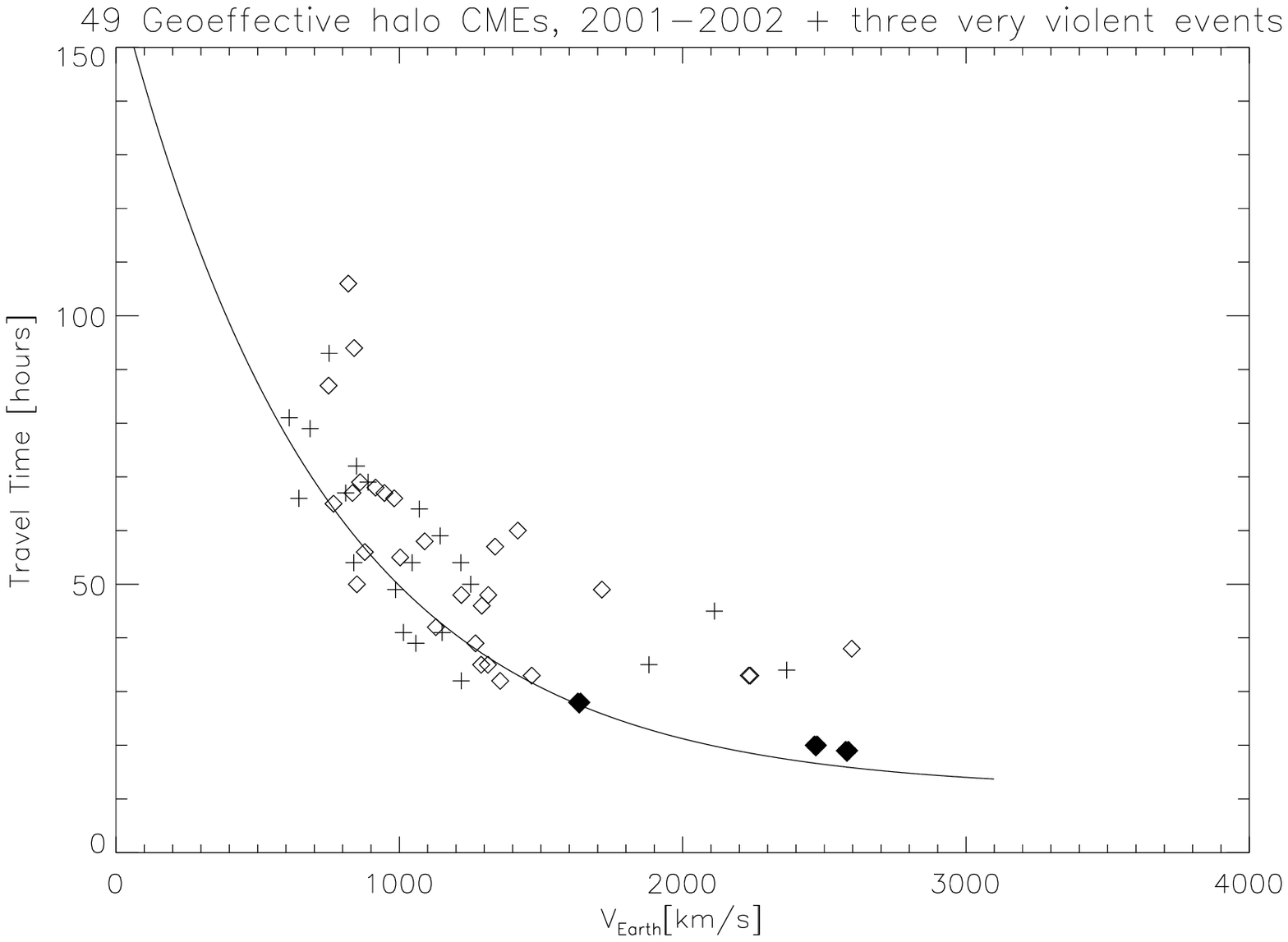}
%\end{subfigure}
\caption{The scatter plots of the space  (left panel) and Earth
directed velocities versus the HCME travel time (TT). Diamond and
cross symbols represent events originating from the western and
eastern hemispheres, respectively. The dot-dashed line (left panel)
is a polynomial fit to all the data points. The continuous line
(right panel) is the ESA model representation. The three additional
dark diamonds (only on the right panel) show the HCMEs (2000/07/14,
2003/10/28 and 2003/10/29) of historical importance.}
\end{figure*}

%\caption{The scatter plot of the space velocities versus the HCME
%travel time (TT). Diamond and cross symbols represent events
%originating from the western and  eastern hemispheres, respectively.
%The dot-dashed line is a polynomial fit to all the data points.}
%\label{time1}
%\end{center}
%\end{figure}

\subsection{Magnitudes of geomagnetic storms}
Magnitudes of geomagnetic disturbances depend not only on the
velocities of CMEs but also on the location of source region on the
solar disk (\emph{e.g.} Gopalswamy, Yashiro, and Akiyama, 2007).
 For our cone model positions of the source regions are characterized by the parameter
$\gamma$, which is the angular distance of the CME from the plane of
the sky. This parameter decides which part of a HCME hits  Earth.
Events with small $\gamma$ strike  Earth with their flanks while
those with large   $\gamma$ hit  Earth with their central  parts.
Figure~3 shows the scatter plot of the plane of the  sky  speeds
multiplied by $\gamma$ versus $D_{ST}$ index. The parameter $\gamma$
was determined from the location of the associated flares. There is
a slight correlation between $V*\gamma$ and $D_{ST}$. Correlation
coefficients are: $\sim$0.49 for the western and $\sim$0.30 for the
eastern events, respectively.
\begin{figure}    %%%%%%%%%%%%%%%%%% FIGURE 1
\begin{center}
\hbox{
\psfig{file=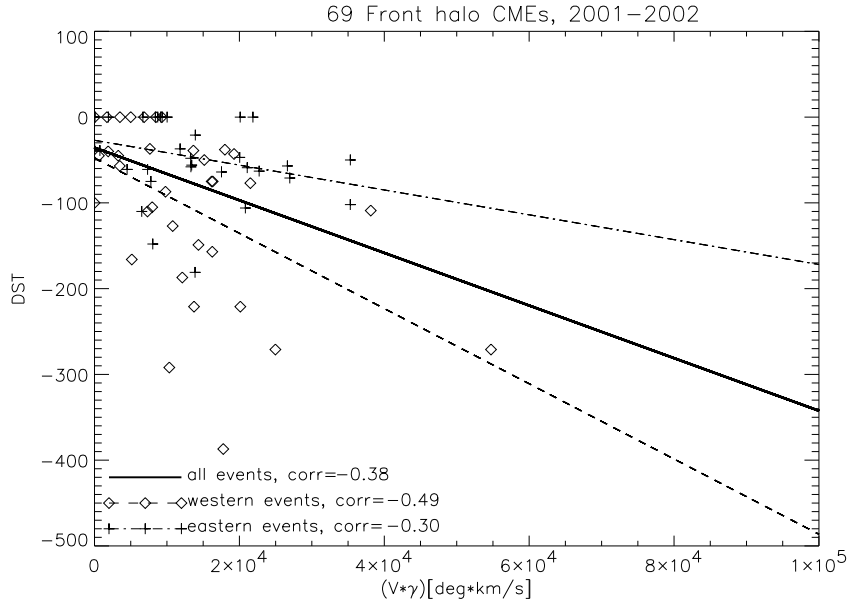,width=11.5cm,clip=}}

\caption{The scatter plot of the sky-plane   speeds multiplied by
$\gamma$ versus $D_{ST}$ index. Diamond  and cross symbols represent
events originating from the western and eastern hemispheres,
respectively. The solid line is a linear fit to  all the data
points, the dot-dashed line is a linear fit to the eastern events,
and the dashed line is a linear fit to the western events.}
\label{time1}
\end{center}
\end{figure}

For comparison, Figure~4 show $V*\gamma$  plot but for the space
parameters. Now the  parameters ($V_{space}$,$\gamma$) were
estimated from the model  (see Michalek
 2006a). From the
inspection of the figure it is clear that the correlation between
($V_ {space}*\gamma$) and $D_{ST}$ is more significant. Correlation
coefficients are:  $\sim$0.85 for the western and $\sim$0.58 for the
eastern events, respectively.
\begin{figure}    %%%%%%%%%%%%%%%%%% FIGURE 1
\begin{center}
\hbox{
\psfig{file=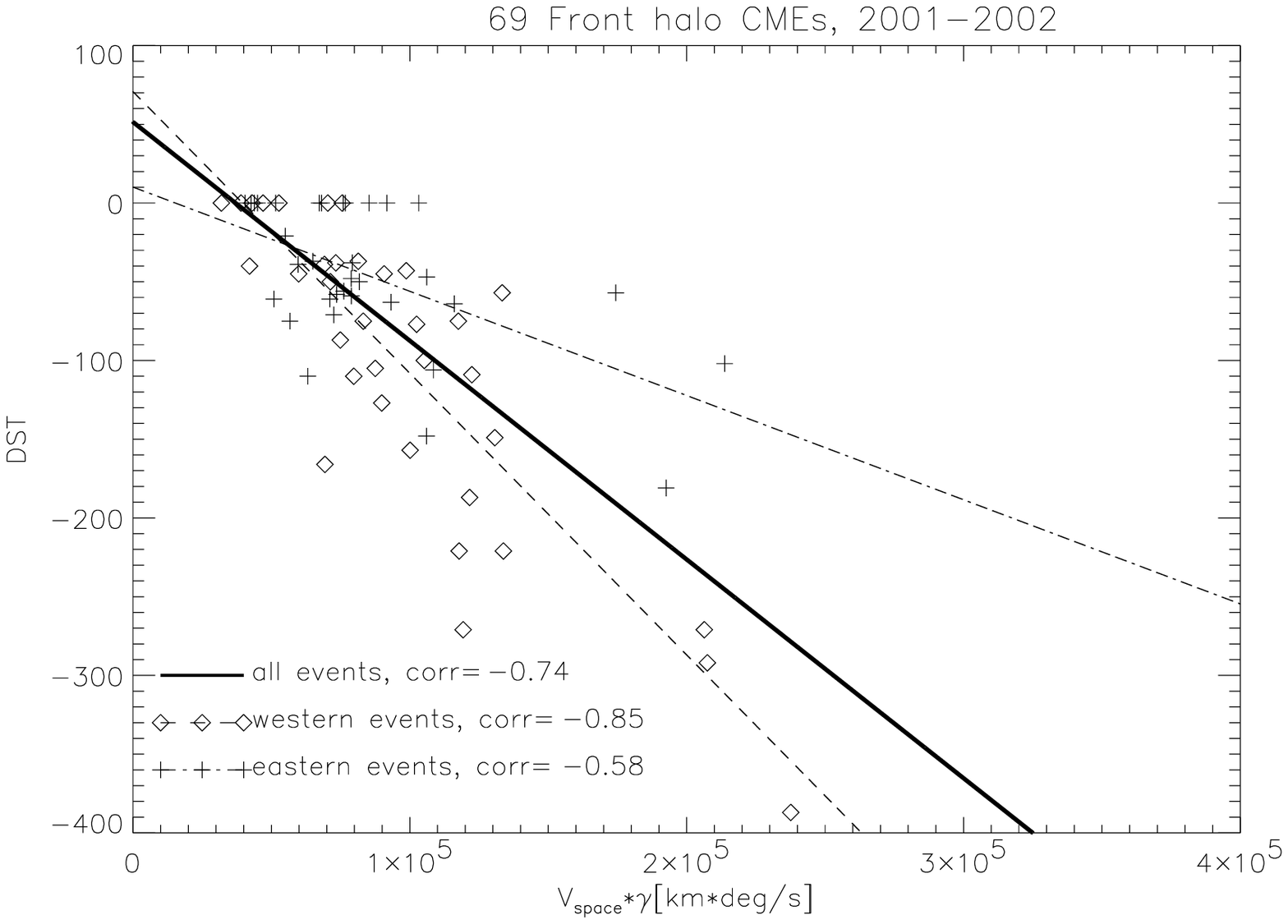,width=11.5cm,clip=}}

\caption{The scatter plot of $V_{space}*\gamma$
 versus $D_{ST}$ index. Diamond symbols represent events
originating from the western and eastern hemispheres, respectively.
The solid line is a linear fit to all the data points, the
dot-dashed line is a linear fit to the eastern events, and the
dashed line is a linear fit to the western events.} \label{time1}
\end{center}
\end{figure}
It is clear that the space parameters, determined from the
asymmetric cone model, could be very useful for space weather
applications. Correlation coefficients are almost two times larger
than those obtained from the projected speeds. For these plots
(Figure~3 and Figure~4), we used all HCMEs from Table~1, even the
non-geoeffective ones. These events generate false alarms.
Non-geoeffective HCMEs are slow (V$<$900km~s$^{-1}$) or have source
region closer to the solar limb. The limb HCMEs appear as halo
events only due to compression of pre-existing coronal plasma. The
investigation confirms  that the western events are more
geoeffective than the eastern ones (\emph{e.g.} Zhang {\it et al.},
2003). Our investigation suggests   that the severest geomagnetic
storms (with $D_{ST}<-200nT$) were generated by the western events,
although  east-hemisphere CMEs are capable of causing such kind of
storms as well (Gopalswamy, \emph{et al.}, 2005a; Dal Lago \emph{et
al.}, 2006).
\section{Summary}
The prediction of the magnitudes and onsets of geomagnetic storms is
crucial for space weather forecasting.  Unfortunately, parameters
characterizing HCMEs, due to the projection effect, are poorly
correlated with geomagnetic disturbances. In the present paper, we
applied  the asymmetric cone model (Michalek, 2006a) to obtain space
speeds and source locations of all front-sided HCMEs observed by
SOHO/LASCO in the period of time from the beginning of 2001 until
the end of 2002. These parameters were used for prediction of the
strength ($D_{ST}$) and onsets (TT) of geomagnetic storms (Figure~2
and Figure~4). The results are very promising. Correlation
coefficients between the space speeds and parameters characterizing
geomagnetic storms (TT and $D_{ST}$) are very significant and almost
two times larger in comparison with results for the projected
speeds. The standard error in the prediction of the travel time is
equal to $\sim$10 hours, almost $60\%$ lower than for the projected
speeds. It is interesting to compare ours results to other cone
models. Xie et al. (2006) calculated absolute differences between
predicted (using the ESA model, Gopalswamy \emph{et al.}, 2005b) and
observed shock travel times for the previous cone models (XOL, MGY,
ZPL). They found that the mean errors for those models were: 6.5,
12.8 and 9.2 hours, respectively. In the present considerations, the
mean difference  between predicted (using polynomial fit from
Figure~2) and observed shock travel times is 8.4 hours, four hours
less than in our previous cone model (MGY). Many authors considered
relation between speeds and geoeffectivenes of CMEs [\emph{e.g.}
Tsurutani and Gonzales, 1998; Lindsay \emph{et al.}, 1999; Cane,
Richardson, and St.Cyr, 2000; Wu and Lepping, 2002; Srivastava and
Venkatakrishnan, 2002; Yurchyshyn, Wang, and Abramenko, 2004). Those
studies demonstrated that the initial speeds of CMEs are correlated
with the $D_{ST}$ index but because they applied the plane of the
sky speeds correlation coefficient were not significant.
 Recently, Michalek
\emph{et al.} (2006b) showed  that the correlation between the space
speed of HCMEs and $D_{ST}$ index could be much more significant
(correlation coefficient was $\sim$0.60). In the present study we
considered  the correlation between $V_{space}*\gamma$ and $D_{ST}$
index. We found that this corelation could be very significant ( for
the western events it is $\sim$0.85). This confirms previous results
that geoeffectivness of HCMEs depends not only on the HCMEs speeds
but also on the direction of their propagation (Moon \emph{et al.},
2005, Michalek \emph{et al.}, 2006b; Gopalswamy, Yashiro, and
Akyama, 2007).
 The present study
shows that the asymmetric cone model could be very useful for the
space weather forecast. There are two important advantages of this
method. First, using our asymmetric cone model  can help predict
space weather with good accuracy.
%To obtain the space
%parameters we need only to determine the height-time plots from
%LASCO observations and implement them to the asymmetric cone model
%fitting procedure.
Second, to predict space weather we need observational data from one
instrument only (a coronagraph along the Sun-Earth line such as the
LASCO coronograph). The method has also some limitations. Faint
HCMEs could not be used for this study because it is difficult to
get the height-time plots around the entire occulting disk.
Fortunately such poor events are generally not geoeffective so they
are not of immediate concern (we missed only one front-sided HCME).
We consider  a flat cone model (not an ice-cream cone model) so in
some cases the measured projected velocities, and as a consequence,
the space speeds could be slightly overestimated.  We need to keep
in mind that the magnetic field direction at the front of magnetic
cloud (or ICME) determines to a large degree the geoeffectiveness of
events. Unfortunately this in-situ measurement can only be recorded
at Earth's vicinity and it cannot be used for the space weather
forecasting due to time constraints. When considering the asymmetric
cone model, it is important to note that CMEs have more complicated
3D structures (Cremades and Bothmer, 2004) and more factors need to
be determined to have a better understanding of what produces the
geomagnetic storms at Earth.

%\end{article}
%\end{document}
\begin{acknowledgements}
Work done by Grzegorz Michalek was  supported by {\it MNiSW} through
the grant N203 023 31/3055 and NASA (NNG05GR03G).
\end{acknowledgements}

\begin{table*}
{\normalsize \textbf{ List of frontsided halo CMEs (2001-2002)}}
{\tiny

\begin{tabular}{c c c c c c c c c }

 \hline \hline

DATA & V & Width & $V_{space}$ & error & $\gamma$   &  Source Location &$D_{ST}$& TT    \\

\hline
&  km/s &      deg  &  km/s &       km/s  &   deg  &    &  nT & hours  \\

\hline

2001/01/10  & 832 & 59 &  1290 & 57 &   80 &   S04E09 &  -&-\\
2001/01/20a & 839 & 111 &   893 &  55 &   57 &   N06E32 &  -61 &66\\
2001/01/20b & 1507 &115 &   1513 & 219 &  47 &   N09E42 & -61 &64\\
2001/01/28 & 916 & 152 &  1080 & 95 &   39 &   S19W48 & -40 &68\\
2001/02/10 & 956 & 59 &  1090 & 63 &   75 &   N14E04 & -50  &69\\
2001/02/11 & 1183& 93 &  1150 & 106 &   62 &   N13W24 & -50  &50\\
%2001/02/15 &  625 & --- & --- & --- &  --- &   ---    & ---  & ---\\
2001/03/19 & 389 & 54 &   700 & 33 &   81 &   N01E08 & -75  &81\\
2001/03/24 &  906 & 71 &  1088 & 88 &   70 &   N15E13 &  -56  &72\\
2001/03/25 &  677 & 81 &  1070 & 115 &   70 &   N16W12 &  -87  &67\\
2001/03/28 & 519 & 134 &   540 & 59 &   75 &   S13E07 &   -  &-\\
2001/03/29 &  942 & 53 &  2731 & 139 &  87 &   N02W02 &  -387 &38\\
2001/04/01 & 1475 & 96 &   1470 & 85 &   58 &   N04E31 & - &-\\
2001/04/05 & 1390 & 115 &  1360 & 117 &   58 &   N13E29 & -59  & 49\\
2001/04/06 & 1370 & 141 &  1243 & 116 &   75 &   N07E13 & -63 &41\\
2001/04/09 &  1192 & 76 &  1549 & 54 &   77 &   S08W10 & -271 &46\\
2001/04/10 &  2411 & 81 &  2680 & 161 &   77 &   S11W06 & -271 &33\\
2001/04/11 &  1103 & 66 &  1423 & 95 &   72 &   S12W12 &  -77 &42\\
2001/04/12 & 1184 &  70 &  1610 & 98 &   73 &   S04W16 & -75 &35\\
2001/04/26 & 1006&  62 &  1396 & 55 &   76 &   N12E09 & -47  &41\\
2001/08/14 &  618 & 62 &  1042 & 70 &   84 &   N05W02 & -105 &67\\
2001/08/25 & 1529 & 141 &  1529 & 144 &   44 &   S27E38 &   -  & -\\
2001/09/11 & 791 & 105 &   803 & 79 &   56 &   N03E33 &  - &-\\
2001/09/24 &  2402 & 68 &  3010  &115 &   71 &   S09E16 & -102 & 34\\
2001/09/28 &  846 & 65 &  1293 & 33 &   82 &   S08E01 & -148 & 59\\
2001/10/01 & 1405 & 111 &  1415 & 175 &   49 &   S30W29 & -166 & 55\\
2001/10/09 & 973 & 101 &  1116 & 51 &   65 &   S25E02 &  -71  &54\\
2001/10/19 & 558 & 69 &   803 & 47 &   69 &   N01W21  &  -  &-\\
2001/10/19 & 901 & 63 &  1465 & 26 &   83 &   N03W06 & -187 &48\\
2001/10/22 & 1336 & 51 &  2180 & 76 &    80 &   S05E08 & -57  &35\\
2001/10/25 & 1090 & 76 &  1335 & 50 &   75 &   S14W02 & -157  &58\\
2001/11/01 & 453 & 57 &   732 & 52 &   72 &   N06W16 & -  &-\\
2001/11/03 & 457 & 128 &   560 & 41 &   57 &   N29W15 & - & -\\
2001/11/04 & 1810& 74 &   2530 & 108 &  82 &   N01W07 & -292  &33\\
2001/11/17 & 1379 & 130 &  1460 & 112 &   54 &   N25E26 & -48  &54\\
2001/11/21 & 518 & 92 &   615 & 32 &   70 &   S15W13 &   - & -\\
2001/11/22a & 1443 & 70 &  1683&  71  &  70 &   S15W17 & -221 &35\\
2001/11/22b & 1437 & 100 &  1833 & 82  &  73 &   N06W16 & -221 &33\\
2001/11/28 & 500 & 75 &   850 & 30 &   76 &   S10E09 &   - &  -\\
2001/12/13 & 864 & 104 &   910 & 49 &   76 &   N13W05 & -39  &87\\
2001/12/14 & 1506 & 130 &  1493 & 143 &  43 &   S22E42 & -39 & 39\\
2001/12/28 & 2216 & 131 &  2073 & 164 &  43 &   S28E52 & -  & - \\
2002/01/04 & 896 &131 &  1096  &160 &   40 &   N41E30 & - &-\\
2002/01/14 & 1492 &138 &  1600 & 109 &   55 &   S15W31 &  - & -\\
2002/02/20 & 952 & 77 &  965 & 90  &  65 &   S02W24 &  - & -\\
2002/03/10 & 1429 & 93 & 1475 & 137  &  62 &   S09E26 &  - & -\\
2002/03/11 & 950 & 107 &  955 & 34  &  54 &   S18E31 &  - & -\\
2002/03/14 &  961 &99 & 1000 & 46  &  65 &   S25E01 & -37  &93\\
2002/03/15 & 957 & 114 & 1030 & 47  &  79 &   N10W02 & -37 &62\\
2002/03/18 & 989 & 101 & 947 & 42  &  64 &   N11W22 &  - &  -\\
2002/03/22 & 1750 & 81 & 1725 & 149  &  61 &   S11W27 & -100  &39\\
2002/04/15 & 720 & 80 & 1033  &31  &   87 &   S01W02  &-127 &56\\
2002/04/17 & 1240& 61 & 1720 & 48  &  76 &   N06W12  &-149 &60\\
2002/04/21 & 2393 &100 & 2381 & 325 &   56 &   S10W32 & -57  &49\\
2002/05/07 & 720 & 68 &  831 & 39  &  76 &   S05E13  &-110 &79\\
2002/05/08 & 614 & 50 &  961 & 56  &  83 &   S05W05  &-110  &69\\
2002/05/16 & 600 & 60 & 1022 & 38  &  77 &   S10E08 & -58  &67\\
2002/05/22 & 1557 & 68 & 1724 & 86  &  71 &   S12W14  &-109  &32\\
2002/07/15 & 1151 & 67 & 1081 & 28  &  69 &   N18E10 & - & -\\
2002/07/18 & 1099&  104 & 1110 & 66  &  66 &   N20W13 & -38 & 94\\
2002/07/20 & 1941 & 125 & 1683  &255  &  46  &  N01E44 & -38 & 32\\
2002/07/23 & 2285 &106 &  2018  &328  &  81 &   S08E29 &  - &  -\\
2002/07/26 & 818 & 80 &  846 & 36  &  65 &   S21E13  &-& -\\
2002/08/16 & 1585 & 88 & 1576 & 68  &  74 &   S14E06  &-106 & 54\\
2002/08/22 & 998 &119 & 1151 & 133  &  52 &   S27W27  &-45 &106\\
2002/08/24 & 1913 &117 & 1890 & 217  &  48 &   S20W37  &-45 & 57\\
2002/09/05 & 1748 &41 & 2638  &52  &  75 &   S08E12  &-181 & 45\\
2002/11/09 & 1838 & 92 & 1673  &96  &  63 &   S16W22  &-43 & 48\\
2002/11/24 & 1077 &70 &  1433 & 70  &  81 &   N06E06  &-64 &50\\
2002/12/19  & 1092 &75 &   1155 & 49  &  72 &   N06W17  &-75  &68\\

\hline
\end{tabular}
}
\end{table*}

\end{article}
\end{document}